\documentclass[pre,showpacs]{revtex4}
\usepackage{graphics}  
\newcommand{\marge}[1]{\marginpar{}}  % do not show margin notes
\newcommand{\Sl}[1]{{}}           % do not show labels
\newcommand{\beq}[1]{\Sl{#1}\begin{equation}\if#1\empty\else\label{#1}\fi}
\newcommand{\eeq}{\end{equation}}
\newcommand{\beqa}[1]{\Sl{#1}\begin{eqnarray}\if#1\empty\else\label{#1}\fi}
\newcommand{\eeqa}{\end{eqnarray}}

\newcommand{\nm}{\nonumber\\}
\newcommand{\Eq}[1]{(\ref{#1})}

\newcommand{\la}{\langle}
\newcommand{\ra}{\rangle}
\newcommand{\im}{\imath}

\begin{document}

\title{Propagation-Dispersion Equation} 
\author{Jean Pierre Boon}
\email{jpboon@ulb.ac.be}
\homepage{http://poseidon.ulb.ac.be/} 
\author{Patrick Grosfils}
\email{pgrosfi@ulb.ac.be} 
\author{James F. Lutsko}
\email{jlutsko@ulb.ac.be} 
\affiliation{Center for Nonlinear Phenomena and Complex Systems\\ 
Universit\'{e} Libre de Bruxelles, 1050 - Bruxelles, Belgium}
\date{\today}

\begin{abstract}

A {\em propagation-dispersion equation} is derived for the first
passage distribution function of a particle moving on a substrate 
with time delays. The equation is obtained as the continuous limit 
of the {\em first visit equation}, an exact microscopic finite 
difference equation describing the motion of a particle on 
a lattice whose sites operate as {\em time-delayers}. 
The propagation-dispersion equation should be contrasted with the
advection-diffusion equation (or the classical Fokker-Planck equation) 
as it describes a dispersion process in {\em time} (instead of diffusion
in space) with a drift expressed by a propagation speed with non-zero 
bounded values. The {\em temporal dispersion} coefficient is 
shown to exhibit a form analogous to Taylor's dispersivity. 
Physical systems where the propagation-dispersion equation 
applies are discussed.


\end{abstract}

\maketitle

\section{Introduction}
\label{intro}

Often to describe the microscopic mechanism of a diffusion process, 
one considers a test particle executing a random walk on some substrate, 
and one writes a mean-field equation in terms of the probabilities that 
the particle performs elementary displacements in given or arbitrary 
directions. The question one then asks is: 
where will the particle be after some given time (in the long-time limit)? 
The answer is given by the distribution function $F(r,t)$, the probability 
that, given the particle was initially at $r=0$ at $t=0$, it will
be at position $r$ at time $t$ 
(for $t \rightarrow \infty$, that is for $t$ large compared to the 
duration of an elementary displacement). $F(r,t)$ is obtained as the 
solution to the Fokker-Planck equation for diffusion, and one finds
that, in the the long-time limit, $F(r,t)$ is Gaussianly distributed in 
space \cite{feller}.

When there is an interactive process between the particle and the 
substrate such that the particle undergoes directed motion subjected 
to time delays, it is interesting to view the motion in terms of
first passages, and one expects that the long-time dynamics will 
be different from that described by the usual diffusion equation. 
We obtain indeed a new equation for the long-time behavior of the 
{\em first visit} distribution function $f(r,t)$ of a 
particle whose dynamics is governed by a distribution of time delays. 
The main results in this paper are (i) the {\em propagation-dispersion 
equation}
\begin{eqnarray*}{}
\frac{\partial}{\partial r}f(r,t)\,+\,\frac{1}{c}\,\frac{\partial}
{\partial t} f(r,t)\,=\,
\frac{\gamma}{2}\,\frac{\partial^2}{\partial {t^2}} f(r,t)\;,
\end{eqnarray*}
where $c$ is the propagation speed, and (ii) the expression for the 
{\em time dispersion coefficient} in terms of the covariance of the
reciprocal velocity fluctuations
\begin{eqnarray*}{}
\gamma \,=\,\xi\,\left( \langle \frac{1}{v^2} \rangle - 
\frac{1}{c^2} \right)\,,
\end{eqnarray*}
where $\xi$ is a characteristic correlation length. We first give
a heuristic derivation of the propagation-dispersion equation using
multi-scale analysis which is then further substantiated by a more
mathematically rigorous development. Using the latter method, we show
that the above results generalize to inhomogeneous systems with
$c \rightarrow c(r)$ and $\gamma \rightarrow \gamma (r)$.

A characteristic example where the equation applies is particle 
dispersion in a granular medium as studied experimentally by Ippolito 
{\em et al.} \cite{hulin}, as we shall discuss below. 
On the other hand, there are prototypical abstract systems, for
particle-substrate interactive dynamics, 
such as the automaton known as Langton's ant \cite{ant} and other 
simple automata \cite{cohen,grosfils}, where particle-substrate 
interactions produce time delays in the dynamics. 
These automaton systems offer
the advantage of explicit microscopic dynamics which can be
solved exactly. For instance Grosfils, Boon, Cohen, and Bunimovich 
\cite{grosfils} developed a one-dimensional automaton for 
which they provided a mathematical analysis also applicable to the 
two-dimensional triangular lattice. 

The one-dimensional case is particularly simple to describe. 
The automaton universe is the one-dimensional lattice where, 
at each time step, a particle moves from site to site, in the 
direction given by an indicator. One may think of the indicator as 
a `spin' ($\uparrow$ or $\downarrow$) defining the state of the site: 
when the particle arrives at a site with spin up, 
it moves to the next neighboring site in the direction of its incoming 
velocity vector, whereas its velocity is reversed if the spin is down.
But the particle modifies the state of the visited site ($\uparrow 
\Longleftrightarrow \downarrow$) so that on its next visit, 
the particle is deflected in the direction opposite to the scattering 
direction of its former visit. With this specific microscopic dynamics,
back-scattering produces time delay. Grosfils {\em et al.} derived  
the equations describing the microscopic dynamics of the particle on the
one-dimensional lattice (and also on the triangular lattice) under 
the general condition that the spins at the initial time can be 
arbitrarily distributed. They proved that the particle will always go 
into a propagation phase, regardless of the initial distribution of 
spins; they also showed that the basic mechanism for propagation is a 
blocking process \cite{grosfils}.
 
It was shown by Boon \cite{boon} that the propagation equations in 
\cite{grosfils} are particular cases of a general equation describing 
the {\em first visit} of the particle to a new site in the propagation 
phase. This equation, obtained in the context of specific lattice 
dynamics, is a first passage equation which has more general 
applicability, and will be our starting point in the present work.

\section{First visit equation}
\label{visit_eq}

Consider a particle propagating in a D-dimensional channel; we project 
its motion onto a one-dimensional lattice - {\em the propagation 
line} - whose sites are labeled by integers $l=0,\,1,\,2,\,...\,$.  
The distance, in lattice units, between neighboring sites on the 
propagation line is denoted by $\delta r$. 
We set the clock at $t_0=0$ when the particle is at site $l=0$ where 
it enters the propagation channel. Its trajectory will intercept 
successively sites $l=1,\,2,\,3,\,...$ for the first time at times 
$t_1,\,t_2,\,t_3\,...$ respectively. The $t_i$'s are integer multiples 
of the automaton time step $\delta t$. While sites $0,\,1,\,2,\,3,\,...$ 
are equally spaced, the time differences between first visits, 
$t_{i+1} - t_i$, are (in general) not equally distributed.

A given (arbitrary) spin configuration defines a set of first
passage times $t_1,\,t_2,\,...\,$. Let $i_r$ be the random variable which 
corresponds to the number of steps required for the particle to reach 
position  $r=l_r\delta r$ for the first time. 
We define $f(r,t)$ as the probability of finding the particle at position 
$r$ for the first time at time $t=t_r=i_r\delta t$; 
$f(r,t)$ obeys the {\em first visit equation}, i.e. the finite 
difference equation \cite{boon}
\beq{a0}
f(r,t)=\sum_{j=0}^n\,p_j\,f(r - \delta r,t-\tau_j)\;.
%\;\;\;\mbox{with}\;\;\; \tau_j\,=\,(1+\alpha j)\,m\,\delta t\;.
\eeq 
Here $p_j$ is the probability that the particle propagates from 
$r - \delta r$ to $r$ in $\tau_j/\delta t$ time steps, i.e. $\tau_j$ is 
the time between two successive first visits on the propagation line. 
The sum is over all possible time delays, weighted by the 
probability~$p_j$ 
%(a polynomial function of $q$, the probability that 
%a site be in the $\uparrow$ state at the initial time)  
with $\sum_{j=0}^n\,p_j\,=\,1$. For specific lattice 
dynamics, this is a condition for the existence of a blocking mechanism 
(described in \cite{grosfils}) responsible for propagation, and an explicit 
expression can then be given for $\tau_j$ (see Appendix).
Here we present a general derivation for which this specification is not 
necessary.

An alternative equivalent formulation of Eq.\Eq{a0} reads
\beq{a0a}
f(r,t)=\sum_{j=0}^\infty\,{\tilde p}_j\,f(r - \delta r,t-j\delta t)\;,
\eeq
which we shall use in Section \ref{gen_fct}. The difference
between the two formulations is that the structure of the distribution
of the delays is contained either in the $\tau_j$'s (Eq.\Eq{a0}) or
in the ${\tilde p}_j$'s (Eq.\Eq{a0a}).
 
Equation \Eq{a0} - or equivalently \Eq{a0a} - expresses the probability 
that the particle be for the first time at position $r$ at time $t$ in 
terms of the probability that it was visiting site $r - \delta r$ at 
earlier time $t - \tau_j$, where $j = 0,1,2,...,n$. 
Given a site $r$ on the propagation line, the particle will infallibly 
reach that site in the course of its displacements; the question then is: 
{\em when will the particle be at position $r$ when $r$ is large?} 
Since the particle executes $t/\delta t$ displacements to cover the distance 
$r$, the answer will be given in terms of the time distribution of the 
probability $f(r,t)$ for large fixed value of $r$, i.e. for $l_r\gg 1$. 
This corresponds to taking the continuous limit of equation \Eq{a0}
in a physically consistent way that we establish in Section \ref{limit}
and in a mathematically rigorous way as shown in Section \ref{gen_fct}.

For one particular realization, the successive time delays are set by
a given spatial configuration of the time delayers, and the time
taken by the particle to perform a displacement from $r-\delta r$ 
to $r$ depends on that configuration. For an ensemble of realizations,
the distribution function of the time delays defines the average 
displacement time
\beqa{a2}
\la \tau \ra\,=\, \sum_{j=0}^n\,\,p_j\,\tau_j
\,=\, \sum_{j=0}^n\,\mu_j\,p_j\,\delta t
\,=\,\la\mu\ra\,\delta t \;,
\eeqa
and the variance
\beqa{a3}
\la \tau^2\ra -\la \tau\ra ^2
&=&\left\{\sum_{j=0}^n \mu_j^2\,p_j\,-[\sum_{j=0}^n \mu_j\,p_j]^2\right\}\,
(\delta t)^2 \nm
&=&(\la \mu^2\ra -\la \mu\ra ^2)\,(\delta t)^2 \;,
\eeqa
where $\mu_j=\tau_j/\delta t$ is the number of time steps during the 
time delay $\tau_j$. The general condition on the $p_j$ distribution 
is that its moments be finite (for specific lattice dynamics, such as
described in the Appendix, they are finite  by virtue of the blocking 
mechanism discussed in \cite{grosfils}). Higher order moments 
$\la\tau^a\ra$ are defined similarly. 
%However since $\la\tau^n\ra\sim (\delta t)^n$ ($n>2$), 
%we shall see that they are negligible in the continuous limit.

\section{Multi-scale analysis}
\label{limit}

Here we discuss how the continuous limit should be taken given that
the system exhibits two time scales which correspond to
(i) a propagation process characterized by the average time necessary 
to complete a finite number of displacements $r/\delta r$ 
\beq{a4}
{\rm E}[t_r]\,=\,\la\mu\ra\,r\,\frac{\delta t}{\delta r}\;,
\eeq
and (ii) the dispersion around this average value characterized by
the variance 
\beq{a5}
{\rm Var}[t_r]\,=\,(\la \mu^2\ra \,-\,\la \mu\ra ^2)\,(\delta t)^2\,
\frac{r}{\delta r}\;.
\eeq
For finite $r$, these are finite quantities. Correspondingly we define
the following quantitites that will be used in the continuous limit 
%$\delta t\rightarrow 0$, $\delta r\rightarrow 0$
\beq{a6}
\frac{1}{c}\,=\,
%\lim_{\delta t\to 0\;,\delta r\to 0}\;
\la\mu\ra\,\frac{\delta t}{\delta r}\;,
\eeq
and
\beq{a7}
\gamma\,=\,
%\lim_{\delta t\to 0\;,\delta r\to 0}\;
(\la\mu^2\ra \,-\,\la\mu\ra ^2)\,\frac{(\delta t)^2}{\delta r}\;.
\eeq
$c$ ($\neq 0$) will be identified as the propagation speed 
(see Section \ref{P-D_Eq}) and $\gamma$ ($\geq 0$) will be identified 
as the dispersion coefficient (see Section \ref{correlations}).
 
We want to compute the continuous limit of the first visit equation
(\ref{a0}), and obtain a partial differential equation for $f(r,t)$.
The procedure must be performed in two successive steps according to 
the scale over which one wants to probe the process when $r$ is large.
This is analogous to multi-scale expansion in the derivation of the 
Fokker-Plank equation \cite{kampen} or of the Navier-Stokes equation 
\cite{multiscale}.
Consider that, to measure first passages, we use a detector with 
tunable resolution. The first visit time $t_r$ goes like $r$ 
(see \Eq{a4}); so in order that the measure be performed with the
same accuracy at any position $r$, we need a resolution
such that $t_r/r$ has always the same order of magnitude.
%therefore $\delta t/\delta r$ must be constant (see \Eq{a6}). 
We can then measure first visit times at any position 
$r$ with a resolution which is appropriate to evaluate the average 
first passage time, i.e. to measure the propagation speed $c$. However, 
when performing measurements over many realizations, the successive 
arrival times at position $r$ (the fluctuations around the average time)
will be poorly resolved (see insets in Fig.1). In order to measure the
dispersion around ${\rm E}[t_r]$ with sufficient accuracy, the detector 
must be adjusted so that dispersion measurements at various positions 
$r$ can be performed with the same resolution. Therefore we impose that 
the width $\Delta T\,=\,\sqrt{{\rm Var}[t_r]}$ of the dispersion curve 
be measured with a resolution $\delta T$ such that the curve always 
contains about the same number of points, i.e. $\Delta T/\delta T$ 
has the same order of magnitude for all positions. Since $\Delta T$ 
grows with the distance like $\sqrt{r}$ (see \Eq{a5}),
$\delta T$ must also go $\sim \sqrt{r}$ in order to obtain an acurate
measure of the dispersion $\gamma$ (see \Eq{a7}).
%therefore $(\delta t)^2/\delta r$ should  be constant.

We may summarize by saying that the continuous limit must be taken 
in a sequential order, first to obtain the average first passage time 
(which gives a measure of the propagation velocity $c$), and second, 
to measure the first passage dispersion in the moving reference frame 
(i.e. around $E[t_r]$). The first step will yield an Euler type equation 
and the second step the propagation-dispersion equation. 
Mathematically, these two steps materialize in the two successive orders, 
${\cal O}(\epsilon^1)$ and ${\cal O}(\epsilon^2)$, of the development in 
Section \ref{P-D_Eq}, where $\epsilon $ is a smallness parameter defined
as $\delta t / t_{obs}$, the ratio of the microscopic time to the
macroscopic observation time. The continuous limit corresponds to
$\epsilon \ll 1$. It follows from the above discussion that there 
are two length scales: (i) the first one corresponds to propagation 
\beq{r_1}
t_{obs}\,c\,=\,t_{obs}\frac{1}{\la \mu \ra}\frac{\delta r}
{\delta t}\,=\, \la \mu \ra^{-1}\epsilon^{-1} \delta r \,,  
\eeq
for which the measurement unit is $\epsilon^{-1} \delta r$;
(ii) the second length scale  corresponds to dispersion
\beq{r_2}
t_{obs}^2\,\frac{1}{\gamma}\,=\,t_{obs}^2\,\frac{1}{\la \mu^2 \ra
- \la \mu \ra^2} \frac{\delta r}{(\delta t)^2}\,=\,(\la \mu^2 \ra
 - \la \mu \ra^2)^{-1}\epsilon^{-2} \delta r \,,  
\eeq
which is thus measured in units $\epsilon^{-2} \delta r$.
It is therefore natural to introduce two space variables, 
$r_1\,=\,\epsilon r$ and $r_2\,=\,\epsilon^2 r$, and correspondingly 
two time variables, $t_1\,=\,\epsilon t$ and $t_2\,=\,\epsilon^2 t$.
Note that $r_1$ and  $r_2$ are merely two expressions of the space
variable $r$ with different scalings, and similarly for $t_1$ and  $t_2$.
Consequently the space and time derivative operators must be rescaled as
\beq{rescale}
\partial_r \rightarrow \epsilon \partial_{r_1} \,+\,\epsilon^2 \partial_{r_2}
\;\;\; ;\;\;\; \partial_t \rightarrow \epsilon \partial_{t_1} \,+\,
\epsilon^2 \partial_{t_2}
\eeq 
Accordingly $f(r,t)$ is expanded as 
$f\,=\,f^{(0)}\,+\,\epsilon f^{(1)}\,+\,\,{\cal O}(\epsilon^2)$, where
$f^{(0)}$ is the distribution function in the absence of dispersion 
($f^{(0)}$ plays the same role as the local equilibrium distribution 
function in the derivation of the Navier-Stokes equation \cite{multiscale}).

While in taking the continuous limit, there is a mathematical separation 
of scales (i.e. one probes either the propagation mode or the dispersion 
mode), in practice, there are 
systems with sufficiently large $\gamma$ (see Section \ref{comments} 
and Appendix) where propagation and dispersion can be measured 
simultaneously, provided one uses a detector with high resolution.

\section{Propagation-dispersion equation}
\label{P-D_Eq}

In order to compute the continuous limit of Eq.\Eq{a0}, 
it is convenient to rewrite the equation as
\beq{a1}
f(r+\delta r,t)=\sum_{j=0}^n\,p_j\,f(r,t-\tau_j)\;,
\eeq 
to which we now apply the multi-scale expansion (in Section \ref{gen_fct}
we give an alternative derivation which is more mathematically rigorous). 

To first order, we obtain
\beqa{epsilon_1}
{\cal O}(\epsilon^1):\;\;\;\; 
\partial_{r_1} f^{(0)}
\,=\,-\frac{\la \tau \ra}{\delta r}\,\partial_{t_1} f^{(0)} 
\,=\,-\la \mu \ra \frac{\delta t}{\delta r} \,\partial_{t_1} f^{(0)}\;,
\eeqa
or, with \Eq{a6}, 
\beq{euler}
\partial_{r_1} f^{(0)}\,+\,\frac{1}{c}\,\partial_{t_1} f^{(0)}\,=\,0\;,
\eeq
which is an Euler equation. 

To second order, we have
\beqa{epsilon_2a}
{\cal O} (\epsilon^2): \;\;\;\;
\partial_{r_2}f^{(0)}
\,+\,\partial_{r_1} f^{(1)}\,+\,\frac{1}{2}\delta r\,
\partial_{r_1}^2 f^{(0)}&=&  \nm
-\,\frac{\la \tau \ra}{\delta r}\,(\partial_{t_2} f^{(0)}\,+\,
\partial_{t_1} f^{(1)})\,&+&\,
\frac{1}{2}\frac{\la \tau^2 \ra}{\delta r}\,\partial_{t_1}^2 f^{(0)}\;,
\eeqa
where we will inject the first order result \Eq{epsilon_1}. This 
substitution can be performed either in the last term on the l.h.s. 
or in the last term on the r.h.s. with two different results of which 
only one can be correct. In fact there is no ambiguity as to which is 
the legal procedure: 
Eq. \Eq{a1} is local in space (not in time) which indicates that in the
continuous limit we must obtain a partial differential equation with
a spatial derivative not exceeding first order. The correct 
substitution of \Eq{epsilon_1} into \Eq{epsilon_2a} gives
\beqa{epsilon_2b}
\partial_{r_2}f^{(0)}\,+\,\partial_{r_1} f^{(1)}\,=\,
-\frac{1}{c}\,(\partial_{t_2} f^{(0)}\,+\,\partial_{t_1} f^{(1)})\,+\,
\frac{1}{2} \left(\frac{\la \tau^2 \ra}
{\delta r}\,-\,\frac{\delta r}{c^2}\right)\,\partial_{t_1}^2 f^{(0)}\;,
\eeqa
where the quantity in parantheses in the second term of the r.h.s. is equal 
to $(\la \mu^2 \ra \,-\, \la \mu \ra^2)(\delta t)^2/{\delta r}\equiv \gamma$.

Summing Eqs.\Eq{euler} and \Eq{epsilon_2b} after multiplication
by $\epsilon$ and by $\epsilon^2$ respectively, we obtain
\beqa{epsilon1+2}
\epsilon\,\partial_{r_1}(f^{(0)}+\epsilon\, f^{(1)})\,
+\epsilon^2\,\partial_{r_2}\,f^{(0)}
=\,-\frac{1}{c}\,\epsilon\,\partial_{t_1}
(f^{(0)}+\epsilon\, f^{(1)})\,
-\,\frac{1}{c}\,\epsilon^2\,\partial_{t_2}\,f^{(0)}
+\frac{\gamma}{2}\,\epsilon^2\,\partial_{t_1}^2\,f^{(0)}\,.
\eeqa
Rearranging terms (incorporating terms of negligible higher order)  
and going back to the original variables yields
\beq{a19}
\partial_r\,f(r,t)\,+\,\frac{1}{c}\,\partial_t f(r,t)\,=\,
\frac{1}{2}\,\gamma\,\partial_t^2 f(r,t)\;,
\eeq
where terms of order $\epsilon > 2$ are omitted.
With the initial condition that at the origin, say at $r=0$,
$f(0,t)\,=\,\delta (t)$, the solution to Eq.\Eq{a19} is 
\beq{a20}
f(r,t)\,=\,\sqrt{\frac{1}{2 \pi}}\,(\gamma\,r)^{-\frac{1}{2}}
\exp \,\left (-\,\frac{(t\,-\,\frac{r}{c})^2}{2\,\gamma\,r}\right )\;.
\eeq
These results are confirmed by the more mathematically rigorous 
derivation given in Section \ref{gen_fct}.
Note that in the case that all $p_j$'s are zero except one 
($\mu_j \equiv \mu$, e.g. for Langton's ant), then $\gamma\,=\,0$, 
and \Eq{a19} reduces to the Euler equation \Eq{euler}
with $c\,=\,{\delta r}/(\mu \,\delta t)$ (see \cite{boon}).

It is clear from \Eq{a19}, that $c$ is a propagation speed, and $\gamma$
is a transport coefficient expressing dispersion in time (instead of
space like in the classical Fokker-Planck equation for diffusion).
Equation \Eq{a19} is the propagation-dispersion equation governing the
first-passage distribution function of a propagating particle subject 
to time delays. Propagation is guaranteed because $c$ is non-zero 
($\la \mu \ra$ is finite); $c$ has a finite maximum value for $n=0$
(i.e. $j=0$ and $p_0=1$), in which case $\gamma =0$ (see \Eq{a3}), 
and there is propagation without dispersion. 
 
In Fig.1 we show that the above analytical results are in perfect 
agreement with the numerical solution of the general equation, 
Eq.\Eq{a0}. We used delay times equally distributed (Fig.1a) and 
exponentially distributed (Fig.1b); explicit values of the 
quantities $c$ and $\gamma$ are given in the figure caption. 
 
\begin{figure}
\resizebox{\textwidth}{!}{\includegraphics{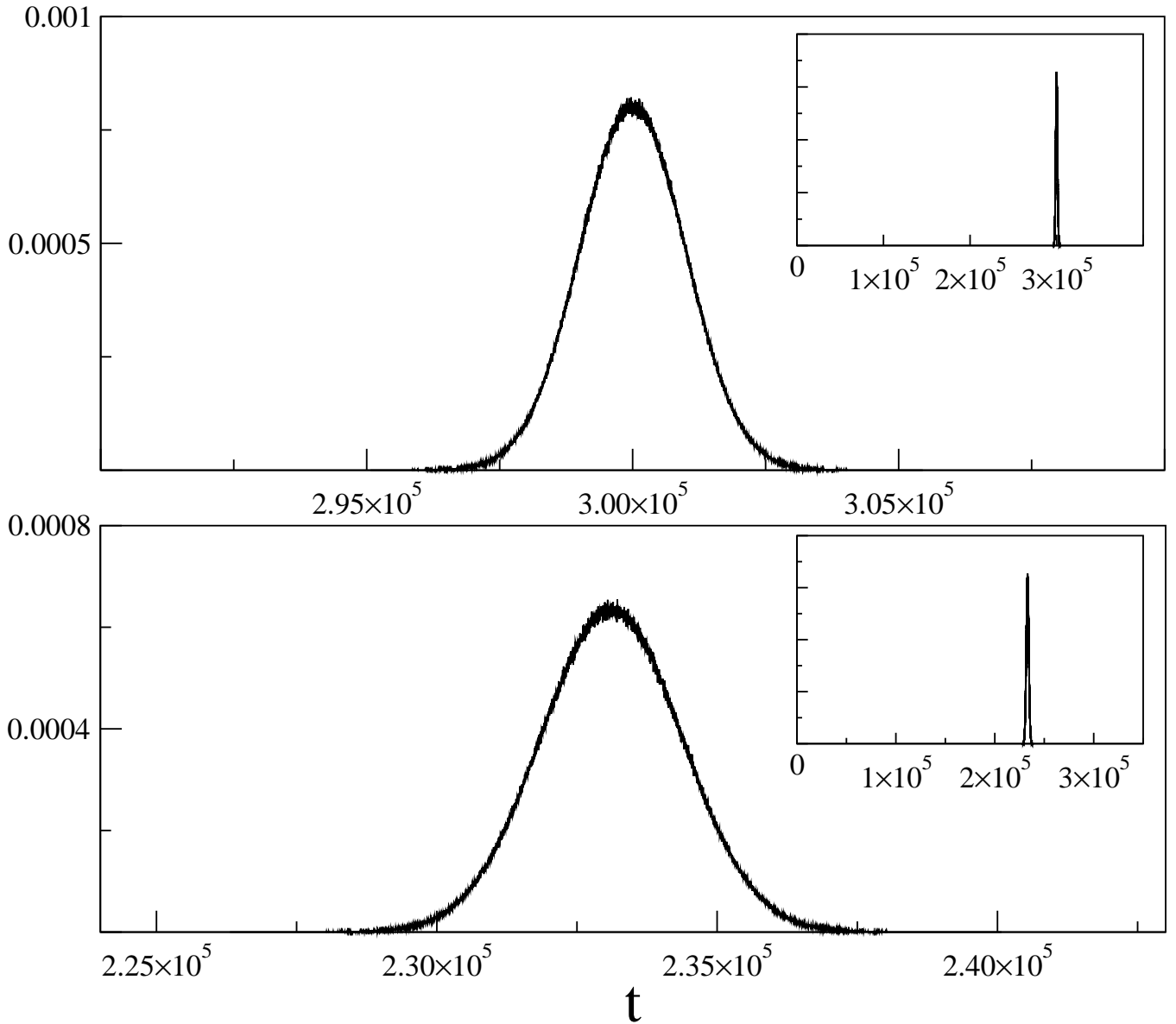}}
\caption{Probability distribution $f(r=3 \times 10^4,\,t)$ 
based on general equation~\Eq{a0}. 
Here $\mu_j\,=\,1 + 2j$, so $c=[(1+2\la j \ra)]^{-1}$.
(a) time delays equally distributed for $j\,=0,\,1,\,\cdots , \,9$, 
with $p_j\,=\,0.1$; $c = 0.1$ and $\gamma = 33$; 
half-width $=\sqrt {2 \gamma r} \simeq 1.41\times 10^3$.
(b) time delays exponentially distributed: 
$p_j\,=\,C \,\exp{-\beta j}$, with $j=0,\,1,\,...,\,9$, $\beta\,=\,0.25$, 
and $C\,=\,[\sum_{j=0}^9 j]^{-1} = 1/45$; $c = 0.128 $ and $\gamma = 52.7$;
half-width $=\sqrt {2 \gamma r} \simeq 1.78 \times 10^3$.
The numerical simulation data and the analytical expression (Eq.\Eq{a20};
solid line, not visible) coincide perfectly. Insets: see text.
In this and subsequent figures, time unit is the automaton time step.}
\end{figure}

\section{Dispersion and correlations}
\label{correlations}

\noindent (i){\em The dispersion coefficient}. 
It follows from Eqs.\Eq{a5} and \Eq{a7}, that $\gamma$ is given by
\beq{b11}
\gamma\,=\,\frac{\la t_r^2 \ra -\,\la t_r \ra^2}{r}\,
\eeq
which, for large $r$, is reminiscent of the classical expression for 
the diffusion coefficient: $D\,=\,\lim_{t\to \infty} \la r^2(t) \ra/2t$. 
Comparison of the two expressions shows interchange of space and time, 
a feature which is illustrated in Figs.2 and 3. In Fig.2, we show three
typical runs on the 1-D spin-lattice by plotting the first passage 
time (minus the mean time of arrival $\la t_r \ra = r/c$) as a function 
of distance $r$. One observes that $[t - \la t_r \ra]$ as a function of
$r$ exhibits fluctuations of the same nature as those obtained 
when plotting the position of a random walker as a function of time. 
In Fig.3 we present simulation data illustrating Eq.(\ref{b11}). 
From measurements performed in the 1-D lattice,  the variance 
$\la t_r^2 \ra-\,\la t_r \ra^2$ is seen to be a linear function of distance 
(see inset) with a slope equal to $\gamma$ (main frame) in the same way
as the diffusion coefficient is obtained as the slope of the mean-square
displacement versus time in the long-time limit of the classical
random walk.  

\begin{figure}
\resizebox{\textwidth}{!}{\includegraphics{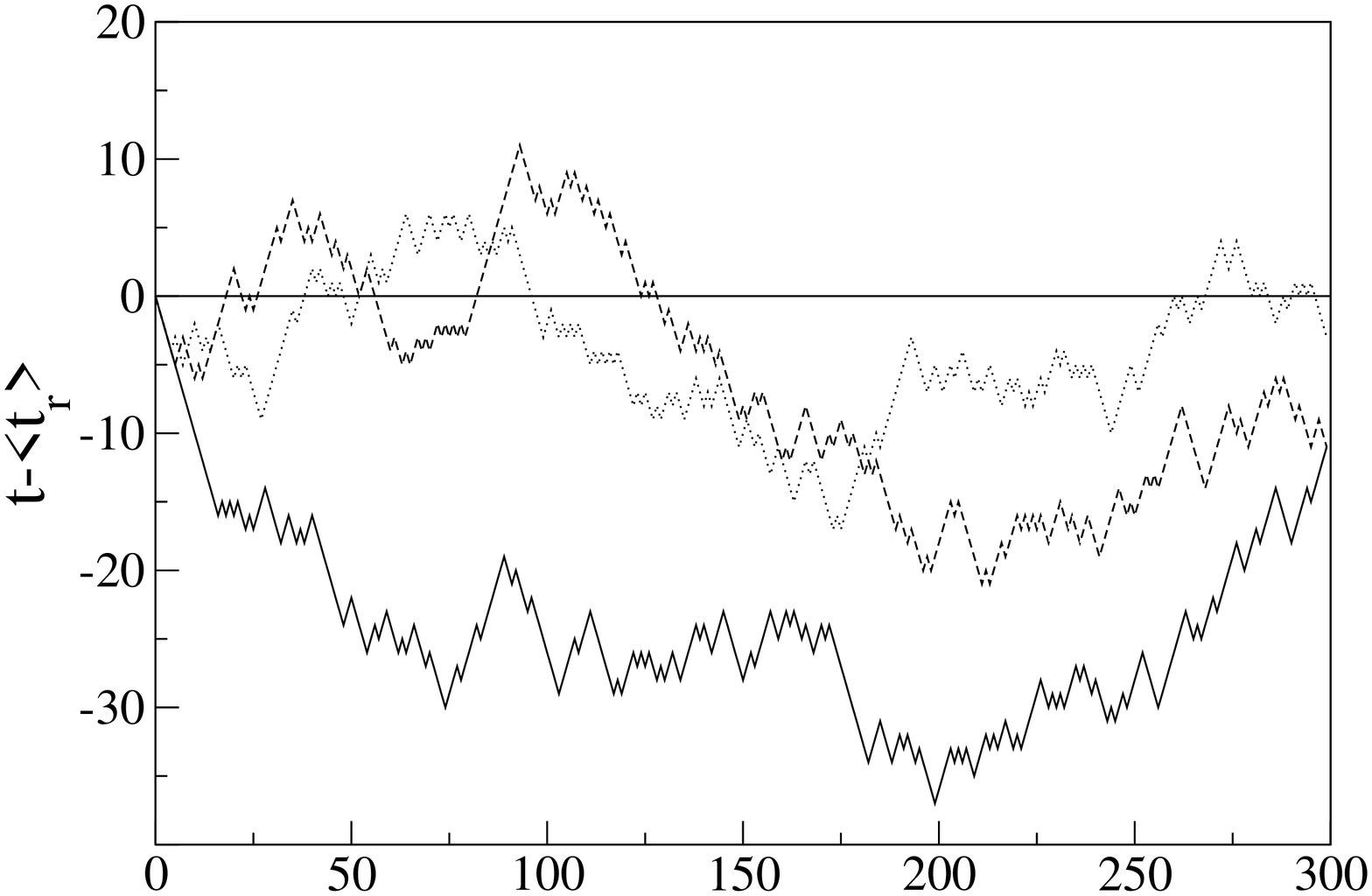}}
\caption{Displacement time 
$(t - \langle t_r \rangle \,=\, t - r/c)$ as a function 
of distance $r$ for three different realizations in the 1-D lattice 
(see Appendix: $q=.5$, and $j=0,1$).
Vertical axis: time in automaton time steps; horizontal
axis: distance in lattice unit lengths.}
\end{figure}

\begin{figure}
\resizebox{\textwidth}{!}{\includegraphics{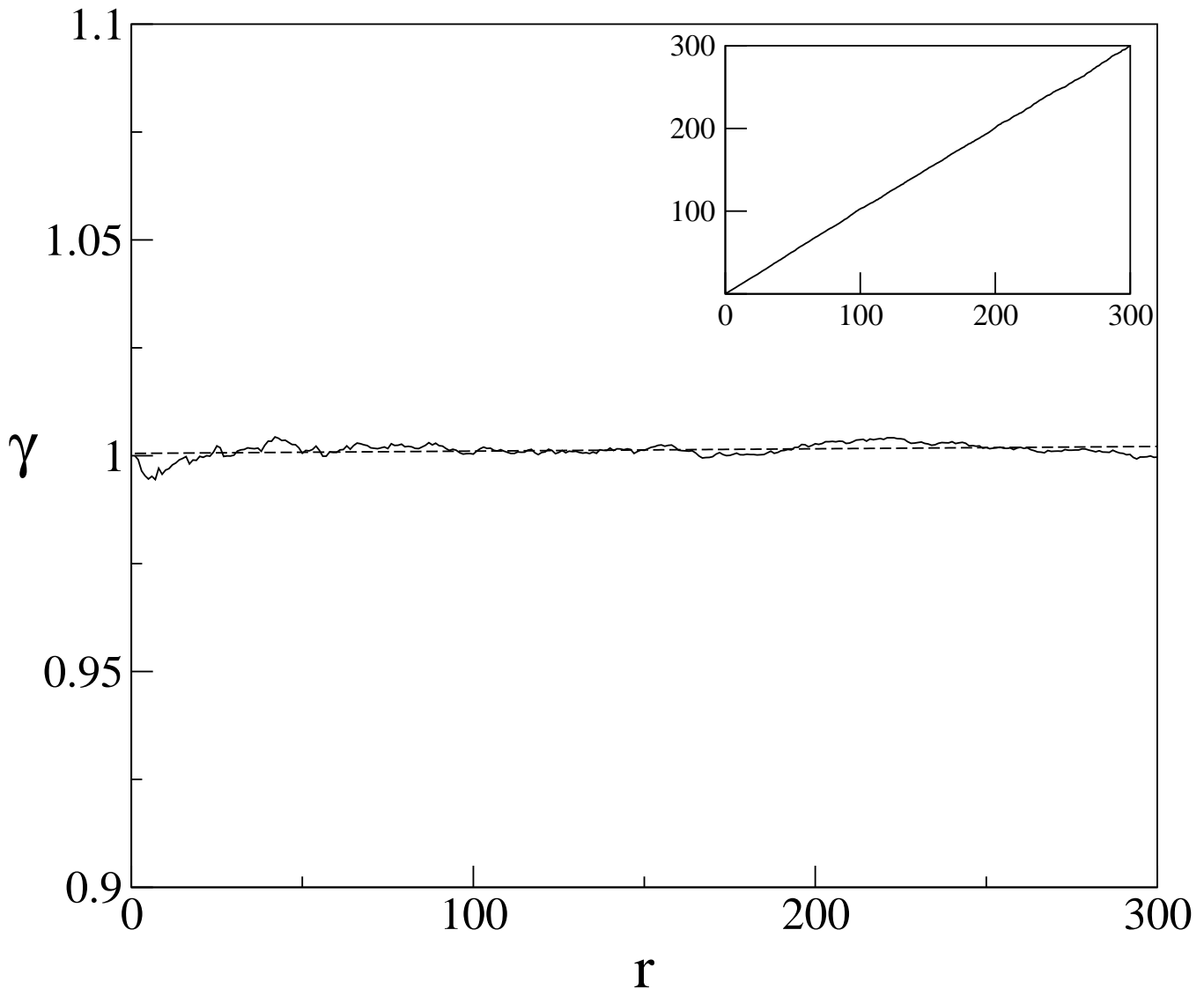}}
\caption{Dispersion coefficient, see Eq.\Eq{b11}, in $1-$D lattice 
(see Appendix: $q=0.5,\, j=0,1,\, p_0=q,\, p_1=1-q,$ and  $\mu_j=1+2j$). 
Inset shows $\la t_r^2\ra\, -\,\la t_r \ra^2$ as a  function of $r$,  
for one single realization, and main frame shows $\gamma$ measured from 
the slope in inset; the horizontal dashed line is the theoretical value: 
$\gamma = (\la\mu^2\ra\,-\,\la \mu \ra^2)=
q+(1-q)3^2\,-\,(3-2q)^2=4q(1-q)=1$. $r$ is in lattice unit lengths and
$\gamma$ is in automaton time steps squared per lattice unit length.}
\end{figure}

\bigskip

\noindent (ii) {\em The correlation function}. 
From Eqs.\Eq{a4} and \Eq{a6}, we have  
$\frac{1}{c}\,=\,\frac{\la t_r \ra}{r}$. 
Here $t_r$ can be expressed in terms 
of the {\em local propagation velocity} $v(r)$, a fluctuating quantity with 
average value $c$. In fact it is the reciprocal local velocity which is
physically relevant: it is the time taken by the particle to 
propagate from position $r$ to $r + \delta r$ (divided by $\delta r$).
Then indeed
\beq{b10}
\la t_r\ra \,=\,\la \int_0^r\,dr'\,\frac{1}{v(r')}\,\ra\,=
\,\int_0^r\,dr'\,\la\frac{1}{v(r')}\,\ra\,=\,\frac{r}{c}\,,
\eeq
which is consistent with the definition of the propagation speed.

Expressing the time $t_r$ in terms of the local velocity 
$v(r)$, we have
\beqa{b12}
\la t_r^2\ra\,=\,\la \int_0^r\,dr'\,\frac{1}{v(r')}\,\int_{0}^{r}\,dr''\,
\frac{1}{v(r'')}\ra
\,=\,\int_0^r\,dr'\,\int_0^r\,dr''\la\frac{1}{v(r')}\,\frac{1}{v(r'')}
\ra\,,
\eeqa
and
\beqa{b13}
\la t_r^2\ra - \la t_r\ra^2 &=&\int_0^r\,dr'\,\int_0^r\,dr''
\la\left(\frac{1}{v(r')}- \la\frac{1}{v}\ra\right)\,
\left(\frac{1}{v(r'')}- \la\frac{1}{v}\ra\right)\ra\,.
\eeqa
In terms of the reciprocal velocity fluctuations
$\delta v^{-1}(r) = v^{-1}(r) - \la v^{-1}\ra =  v^{-1}(r) - c^{-1}$,
\Eq{b13} reads
\beqa{b14}
\la t_r^2\ra - \la t_r\ra^2 &=&\int_0^r\,dr'\,\int_0^r\,dr''
\la \delta v^{-1}(r') \, \delta v^{-1}(r'')\ra\,.
\eeqa
The dynamics of the propagating particle implies that the 
correlation function on the r.h.s. of (\ref{b14}) is
$\delta$-correlated, i.e. 
$\la \delta v^{-1}(r') \, \delta v^{-1}(r'')\ra\,=\,\phi_0\,
\delta (\frac{r'}{\xi}-\frac{r''}{\xi})$ with 
$\phi_0\,=\,\la (\delta v^{-1})^2\ra\,=\,\la \frac{1}{v^2}\ra -
\frac{1}{c^2}$, and where $\xi$ is the elementary correlation length.
So it follows from \Eq{b11} and (\ref{b14}) that 
\beq{c17} 
{\gamma}\,=\,\xi (\langle v^{-2} \rangle - c^{-2})\,,
\eeq 
that is ${\gamma}$ is the covariance of the reciprocal velocity 
fluctuations multiplied by the correlation length (here equal to one 
lattice unit length). This result is analogous to Taylor's
formula of hydrodynamic dispersivity which is expressed as the 
product of the covariance of the velocity fluctuations with a 
characteristic correlation time \cite{taylor}.
We call $\gamma$ the {\em temporal dispersion coefficient}. 
Evidently, there is no dispersion ($\gamma=0$) in the absence of 
velocity fluctuations ($\phi_0\,=\,0$).  

From these results, the explicit computation of the dispersion 
coefficient is straightforward. For instance for the one-dimensional 
spin-lattice (see Appendix; Fig.4a), the local velocity is either $1$, 
with probability $q$, or $1/3$, with probability $(1-q)$, and the 
reciprocal mean velocity is $c^{-1}\,=\,q\,+\,3(1-q)$; so
\beq{b18}
{\gamma}\,=\,q\,+\,3^2(1-q) - c^{-2}\,=\,4q(1-q)\,,
\eeq  
which is exactly the value of $\gamma$ evaluated for the 1-D spin 
lattice in the Appendix and in \cite{grosfils}. A similar 
computation for the triangular lattice (see Fig.4b) yields the
value $\gamma = 72q(1-q)$ in accordance with the result obtained
in \cite{grosfils}.

\bigskip

\noindent (iii) {\em The time current}. Writing Eq.\Eq{a19} as
\beqa{tc1}
(\partial_r \,+\, \frac{1}{c}\,\partial_t)\,f(r,t)\,+\,
\partial_t\,j(r,t)&=&0 \;,
\eeqa
with $j(r,t) = -\frac{\gamma}{2}\,\partial_t\,f(r,t)$, shows formal 
analogy with the classical mass conservation equation
\beqa{tc2}
(\partial_t\,+\,{\bf c} \cdot \nabla)\,\rho({\bf r},t)\,+\,
\nabla\cdot{\bf j}({\bf r},t)&=&0 \;,
%{\bf j}(r,t)&=&-D\,\nabla\,\rho(r,t)\;,
\eeqa
with space and time variables interchanged. So in \Eq{tc1}, $j(r,t)$ 
can be interpreted as a ``current in time ''.

\bigskip  

\noindent (iv) {\em The control parameter}. 
In classical advection-diffusion phenomena, the control parameter is 
the P\'eclet number $P=UL/2D$, where $U$ denotes the mean advection 
speed, $L$, the characteristic macroscopic length, and $D$, 
the diffusion coefficient (see e.g. \cite{koplik}). The analogue for 
propagation-dispersion follows by casting Eq.\Eq{a19} in non-dimensional 
form 
\beq{a13nd}
\partial_{\tt r}\,f(r,t)\,+\,\partial_{\tt t}\,f(r,t)\,=\,B^{-1}\,
\partial_{\tt t}^2\,f(r,t)\;;\;\;\;\; B=\frac{2T}{\gamma c}\;.
\eeq
Here ${\tt r}$ and ${\tt t}$ are the dimensionless space and time
variables: ${\tt r} = r(cT)^{-1}$ and ${\tt t} = t\,T^{-1}$, where
$T$ is a characteristic macroscopic time. $B$ is the control
parameter for propagation-dispersion: it is a measure of the relative
importance of propagation with respect to dispersion. 
Indeed, $B=\frac{2T}{\gamma c}= \frac{2 T^2}{\gamma}\frac{1}{cT}= L_D/L_P$,
i.e. the ratio of the characteristic dispersion length $L_D$ to the
characteristic propagation length $L_P$. At high values of $B$, i.e.
$L_D \gg L_P$, the distribution function is very narrow, and transport
over large distances ($r \geq cT$) is dominated by propagation. 

\bigskip 

\noindent (v) {\em The power spectrum}. 
The propagation-dispersion equation \Eq{a19}, subject to 
the initial condition $f(r=0,t)\,=\,\delta (t)$, describes an initial value 
problem with initial value fixed in space; so we can solve the equation 
by spatial Laplace transformation. Using $\kappa$ as the conjugate space
variable, we obtain
\beq{a24}
\kappa \,{\tilde f}(\kappa ,\omega)\,-\,f(r=0,\omega)\,=\,-\frac{\im\,\omega}
{c}\,{\tilde f}(\kappa ,\omega)\,-\,\frac{\gamma}{2}\,\omega^2\,{\tilde f}
(\kappa ,\omega)\;,
\eeq
where $f(\omega)$ denotes the time Fourier transform. With the initial
condition $f(r=0,t)=\delta(t)$, i.e. $f(r=0,\omega)=\frac{1}{2\pi}$, 
\Eq{a24} yields 
\beq{a25}
{\tilde f}(\kappa ,\omega)\,=\,\frac{1}{2\pi}\,(\,\kappa\,+\,\,\frac{\im\,
\omega}{c}\,+\,\frac{\gamma}{2}\,\omega^2\,)^{-1}\;.
\eeq
This result shows that the system dynamics is  described by one single mode: 
$\kappa\,=\,-\,\frac{\im\,\omega}{c}\,-\,\frac{\gamma}{2}\,\omega^2$. 
The corresponding spectrum $S(k,\omega)\,=\,2\,Re\,{\tilde f}
(\kappa =\im\,k,\omega)$
is given by  
\beq{a26}
S(k,\omega)\,=\,\frac{1}{2\pi}\,\frac{\gamma\,\omega^2}{(k\,+
\,\omega\,/c\,)^2\,+\,\frac{1}{4}\,(\gamma\,\omega^2\,)^2}\;,
\eeq
with $\int_0^{\infty}S(k,\omega)\,dk=1$. The spectrum \Eq{a26} (which  
can also be obtained by double Fourier transformation of \Eq{a20}) 
exhibits a single Lorentzian line typical of a 
diffusive phenomenon, but there are two essential differences with the 
spectrum obtained from the classical advection-diffusion equation: 
(i) the spectrum is a
Lorentzian in $k$ (rather than in $\omega$) with half-width at half-height
$\Delta k = \frac{1}{2}\gamma\omega^2$, indicating that dispersion is 
diffusive in time (instead of space); (ii) the Lorentzian is shifted by a 
quantity proportional to the {\em reciprocal} of the propagation speed.

\section{Exact solution of the first-visit equation}
\label{gen_fct}

In order to emphasize the discrete nature of the problem, we introduce the
notation%
\begin{equation}
g(l,i)=\left( \delta t\right) f(l\delta r,i\delta t), \label{app1}  
\end{equation}%
where here, and below, latin arguments ($l, i, j, \cdots$) always 
indicate integers. The first visit equation, Eq.\Eq{a0a}, is then%
\begin{equation}
g(l,i)=\sum_{j=0}^{\infty}{\tilde p}_{j}^{\left( l-1\right) }g(l-1,i-j)\,.  
\label{app2}
\end{equation}%
Here we consider the general case where the transition probabilities 
$p_{j}^{(l)}$ depend on the lattice position $l$ as indicated
by the superscript (for simplicity we omit the tilde notation).
It is convienient to introduce the generating function
for the distribution which is defined to be%
\begin{equation}
h_{l}(x)\equiv \sum_{i=0}^{\infty }x^{i}g(l,i),  \label{app3}
\end{equation}%
and from which the distribution is obtained via%
\begin{equation}
g(l,i)=\lim_{x\rightarrow 0}\frac{1}{i!}\frac{d^{i}}{dx^{i}}h_{l}(x).
\label{app4}
\end{equation}%
Temporal moments of the distribution can be calculated as%
\begin{equation}
\left\langle j^{a}\right\rangle =\sum_{j=0}^{\infty}j^{a}g\left( l,j\right)
=\lim_{z\rightarrow 0}\frac{d^{a}}{dz^{a}}h_{l}(e^{z})\,,  \label{app5}
\end{equation}%
so that $M_{l}(z)=h_{l}(e^{z})$ is the generating function for moments of
the first passage time at lattice position $l$. The boundary condition that
the particle starts at lattice site $l=0$ at time $i=0$ implies that 
$g(0,i)=\delta _{i0}$, or equivalently $h_{0}(x)=1$.

Substituting Eq.(\ref{app2}) into Eq.(\ref{app3}) gives%
\begin{eqnarray}
h_{l}(x) &\equiv &\sum_{i=0}^{\infty }x^{i}\sum_{j=0}^{\infty}p_{j}^{\left(
l-1\right) }g(l-1,i-j)  \label{app6} \nonumber \\
&=&h_{l-1}(x)\sum_{j=0}^{\infty}p_{j}^{\left( l-1\right) }x^{j},  
\end{eqnarray}%
so that the general solution is%
\begin{equation}
h_{l}(x)\,=\,h_{0}(x)\,\prod_{k=1}^{l}\left( \sum_{j=0}^{\infty}
p_{j}^{\left( k-1 \right)}x^{j}\right)\,.  \label{app7}
\end{equation}%
Using this solution and Eq.(\ref{app5}), exact moments may be easily
calculated; with the normalization $\sum_{j=0}^{\infty}p_{j}^{(k)}=1$, we
obtain%
\begin{eqnarray}
\left\langle j;l\right\rangle  &=&l\,\langle \mu \rangle_{l}\,, 
\label{app8_mu} \\
\left\langle j^{2};l\right\rangle -\left\langle j;l\right\rangle ^{2}
&=&l\sigma _{l}^{2}\,, 
\label{app8_sig} 
\end{eqnarray}%
where%
\begin{eqnarray}
\langle \mu \rangle_{l} &=&\frac{1}{l}\sum_{k=1}^{l}
\sum_{j=0}^{\infty}p_{j}^{\left( k-1\right)}j \,, \label{app9_mu} \\
\sigma _{l}^{2} &=&\frac{1}{l}\sum_{k=1}^{l}\left[ 
\sum_{j=1}^{\infty}p_{j}^{\left( k-1\right) }j^{2}-\left( 
\sum_{j=1}^{\infty}p_{j}^{\left( k-1\right) }j\right)^{2}\right] \,, 
\label{app9_sig}
\end{eqnarray}%
are just averages over the elementary process. In order to develop the
limiting form of the distribution for large $l$, we assume that 
$\langle \mu \rangle_{l}$
and $\sigma _{l}^{2}$ are of order $1$ for all $l$ as is certainly true if the
elementary probabilities are independent of lattice position. We then
introduce a new stochastic variable which measures deviations away from the
expected waiting time as%
\begin{equation}
w_{l}=\frac{\left( j-l\langle \mu \rangle_{l}\right) }
{\sqrt{l\sigma _{l}^{2}}}\,. 
\label{app10}
\end{equation}%
The moment generating function $N_{l}(z)$ for this 
new variable is related to that for the original variable by%
\begin{equation}
N_{l}(z)=\exp \left( -\frac{l\langle \mu \rangle_{l}z}{\sqrt{l\sigma _{l}^{2}}}
\right)\,M_{l}\left( \frac{z}{\sqrt{l\sigma _{l}^{2}}}\right) \,, 
\label{app11}
\end{equation}%
or, using (\ref{app7}) with the boundary condition $h_0(x)=1$,%
\begin{eqnarray}
\ln N_{l}(z) &=&-\frac{l\langle \mu \rangle_{l}z}{\sqrt{l\sigma _{l}^{2}}}+
\sum_{k=1}^{l}\ln
\left( \sum_{j=0}^{\infty}p_{j}^{\left( k-1\right) }
\exp \left( \frac{zj}{\sqrt{%
l\sigma _{l}^{2}}}\right) \right) \,.  \label{app12} 
\end{eqnarray}%
By double expansion of the second term on the r.h.s. of (\ref{app12}),
we obtain
\begin{eqnarray}
\ln N_{l}(z)&\simeq&\frac{1}{2}z^{2}+O\left( \frac{1}{\sqrt{l}}\right) \,, 
\label{app12_exp} 
\end{eqnarray}%

So in the limit of large $l=r/\delta r$, the generating function for the 
moments of $w_{l}$ is just $N_{l}(z)=\exp \left( z^{2}/2\right) $ which 
is recognized as the generating function for a Gaussian distributed 
variable with unit variance. We conclude that in this limit the 
distribution for the original (temporal) variable becomes 
\begin{equation}
g\left( l,j\right)_{l\gg 1}{\,=\,}\frac{1}{\sqrt{2\pi l\sigma _{l}^{2}}}%
\exp \left( -\frac{1}{2}\left( \frac{\left( j-l\langle \mu \rangle_{l}\right)}
{\sqrt{l\sigma _{l}^{2}}}\right)^{2}\right) \,,  \label{app13}
\end{equation}%
from which the distribution for the physical quantities 
(for ${r\gg \delta r}$) reads%
\begin{equation}
f\left( r,t\right){=}
%f\left( r,t\right)_{r\gg \delta r}{=}
\sqrt{\frac{1}{2\pi}}\,\left(\Gamma (r)\,r\right)^{-\frac{1}{2}}
%\frac{1}{\sqrt{2\pi \gamma (r)r}}%
\exp \left( -\frac{\left( t-\frac{1}{C(r)}\,r \right)^{2}}
{2\,\Gamma (r)\,r} \right) \,,  \label{app14}
\end{equation}%
with%
\begin{eqnarray}
\frac{1}{C(r)} \,&\equiv&\,
\frac{1}{C(l\delta r)} = \langle \mu \rangle_{l}\frac{\delta t}{\delta r}
\,=\,\frac{1}{r}\,\sum_{k=1}^{r/\delta r}\langle t \rangle_{k-1}\,,
\label{app15_c} \\
\Gamma(r)  \,&\equiv&\,
\Gamma \left( l\delta r\right)  = \sigma _{l}^{2}
\frac{\left(\delta t\right)^{2}}{\delta r}
\,=\, \frac{1}{r}\,\sum_{k=1}^{r/\delta r}[\langle t^2 \rangle_{k-1}\,-\,
\langle t \rangle^2_{k-1}]\,,
\label{app15_gam}
\end{eqnarray}%
where the definition of $\langle t \rangle_{k-1}$ and 
$\langle t^2 \rangle_{k-1}$ follows straightforwardly from \Eq{app9_mu}
and \Eq{app9_sig}.

This result, \Eq{app14}, is just a realization of the central limit 
theorem. Notice that the development is completely independent 
of any assumption on the magnitude of $\delta r$ and $\delta t$. To make
contact with the dynamical formulation in Section \ref{P-D_Eq} for the 
case of position-independent probabilities, we note that the 
present result is derived with the initial condition $g(0,j)=\delta _{j0}$, 
which means that it is the Green's function for the difference equation. 
In the continuous limit, the corresponding boundary condition is 
$f(0,t)=\delta (t)$ so that the dynamical equation for the distribution 
in this limit should be linear and should have \Eq{app14} as its 
Green's function which implies the 
{\em generalized propagation-dispersion equation}

\begin{eqnarray}
\label{geneq}
\frac{\partial}{\partial r}f(r,t)\,+\,\frac{1}{c(r)}\,\frac{\partial}
{\partial t} f(r,t)\,=\,
\frac{1}{2}\,\gamma(r)\,\frac{\partial^2}{\partial {t^2}} f(r,t)\;,
\end{eqnarray}
with 
\begin{eqnarray}
\label{c(r)}
\frac{1}{c(r)}\,=\,\frac{\langle t \rangle_r}{\delta r}\,
=\,\sum_{j=1}^{\infty}\,p_{j}^{(r)}\,j\,\frac{\delta t}{\delta r}\,,
\end{eqnarray}
and
%\;\;\;\;\;\mbox{and} \;\;\;\;\; 
\begin{eqnarray}
\label{gamma(r)}
\gamma (r) \,=\,[\langle t^2 \rangle_{r}\,-\,
\langle t \rangle^2_{r}]\frac{1}{\delta r}\,=\,
\left[\sum_{j=1}^{\infty}p_{j}^{\left( r\right)}j^{2}-
\left( \sum_{j=1}^{\infty}p_{j}^{\left( r\right) }j\right)^{2}\right] 
\frac{\delta t^2}{\delta r}\,.
\end{eqnarray}
Notice the difference between $c(r)$ and $\gamma(r)$ in \Eq{geneq}
and $C(r)$ and $\Gamma(r)$ in \Eq{app14}: l.c. symbols denote local
quantities whereas capitals indicate space-averaged quantities.
When the waiting time probabilities are space-independent, 
$c(r)\rightarrow c$ and $\gamma (r) \rightarrow \gamma$, and the
generalized equation (\ref{geneq}) reduces to Eq.(\ref{a19}).
An example of microscopic dynamics whose continuous limit is 
described by the generalized equation is discussed in the Appendix.

\section{Comments}
\label{comments}

There is an algebraic similarity in the structure of the 
propagation-dispersion equation (\ref{a19}) and of the classical 
advection-diffusion equation~\cite{feller} which can be formally 
transformed into each other by interchanging space and time variables. 
It should be clear that the two equations describe different, but
complementary aspects of the dynamics of a moving particle.
Solving the propagation-dispersion equation answers the question of 
the time of arrival and of the time distribution around the average
arrival time in a propagation process. It is also legitimate to
ask the complementary question ``where should we expect to find the 
particle after some given time ?'' which should be long compared to 
the elementary time step, but short with respect to the average time 
of arrival. We will then observe spatial dispersion around some
average position which can be evaluated from the solution of the 
advection-diffusion equation. This observation stresses the 
complementarity of the two equations. 

Because the propagation-dispersion equation describes the space-time 
behavior of the {\em first passage} distribution function $f(r, t)$, i.e. 
the probability that a particle be for the first time at some position, 
it describes transport where a first passage mechanism plays an important 
role. So the equation should be applicable to the class of front-type 
propagation phenomena where any location ahead of the front will
necessarily be visited, the question being: {\em when} will a given point
be reached? 

A most interesting case is the
``Diffusion of a single particle in a 3D random packing of spheres''
to quote the title of an article by Ippolito {\em et al.} 
\cite{hulin} where the authors describe an experimental study of
the motion of a particle through an idealized granular medium. 
They measure particulate transport and `dispersivity' which 
corresponds precisely to the quantity $\gamma$ computed
in the present paper. In particular the experimental data 
presented by Ippolito {\em et al.} show  that the mean square 
transit time of the particle through the medium is a linear 
function of the mean transit time (Figs.10 and 11 in \cite{hulin})  
itself a linear function of the percolating distance (Fig.2 in 
\cite{hulin}); this observation is an experimental illustration 
of the features described in the present paper (see e.g. inset
of Fig.5). This experimental study also shows that the particle
transit time is Gaussianly distributed in time (see Fig.9 in
\cite{hulin}) in accordance with the solution (\ref{a20}) of 
Eq.(\ref{a19}) (see Fig.1). 

Front-type dynamics is also encountered in shock propagation
in homogeneous or inhomogeneous media \cite{schock} or packet transport
in the Internet \cite{internet}. As the propagation-dispersion equation 
is for the first-passage time distribution, it should also be suited for 
the description of transport driven by an input current in a disordered 
random medium \cite{kehr}. In the area of traffic flow, there are typical 
situations where cars moving on a highway from location A to location 
B, are subject to time delays along the way, and -- with the assumption 
that all cars arrive at destination --  one wants to evaluate 
the time of arrival \cite{traffic}. Financial series as
in the time evolution of stock values are another example
\cite{financial}: over long periods of time (typically years) one
observes a definite trend of increase of, for instance, the  value
of the dollar. So any preset reachable value will necessarily be attained, 
the questions being: when? and what is the time distribution around the 
average time for the preset value? While the classical question is: 
after such or such period of time, which value can one expect?, there
might be instances where the reciprocal question should be considered.  
Because of the generality of the propagation-dispersion equation, it
should be expected that, either in its simple form \Eq{a19} or in its 
generalized form \Eq{geneq}, the equation will be applicable to a large 
class of first-passage type problems in physics and related domains.

\begin{acknowledgments}

We thank Alberto Su\'arez and Didier Sornette for useful comments. 
JPB acknowledges support by the {\em Fonds National de la Recherche 
Scientifique} (FNRS, Belgium).

\end{acknowledgments}

\section*{Appendix: Specific microscopic dynamics}
\label{dynamics}

The $\tau_j's$ in the first visit equation, Eq.(\ref{a0}), are known
explicitly for specific microscopic dynamics of a particle on a lattice
\cite{boon}. Two parameters are used to characterize the trajectory 
of the particle in the propagation channel. 
First, as the propagation line is not necessarily along one of the lattice 
axes, we define $m \,(\geq 1)$ as the minimum number of time steps required
to travel from site $l$ to the neighboring site $l+1$, i.e. to perform a
displacement $\delta r$ along the propagation line. The second parameter,
$\alpha$, is the length of an `elementary loop', 
i.e. the minimum number of displacements necessary to return to a site. 
Typical values of $m$ and $\alpha$ are shown in Fig.4 for various 
channel geometries. The expression for the time delays then reads
\cite{boon}
\beq{tau}
\tau_j\,\equiv \, \mu_j \,\delta t\,=\,(1+\alpha j)\,m\,\delta t\;.
\eeq

For instance, in the one-dimensional lattice (where $n=1$, $\alpha=2$, 
and $m=1$; see Fig.4a) \cite{grosfils}
$j=0$ corresponds to straightforward 
motion from site $l$ to site $l + 1$ in one time step, and $j=1$ 
corresponds to one step backwards followed by two forward steps.
In the square lattice, $\alpha=4$, and $m=8$ (see Fig.4c) for Langton's 
ant dynamics \cite{ant,boon}. The value $m=8$ follows from the fact 
that when the particle arrives at site 1 for the first time, the shortest 
possible path to the next first visited site on the edge of the 
propagation channel goes to site 2 as shown in Fig.4c. Indeed, for  
Langton's ant dynamics, all sites are initially in the same state 
(here scattering of the particle to the left). So when the particle 
visits site 1 for the first time, the site located immediately North 
of site 1 has not yet been visited and is therefore in the 
left-scattering state. It is then easy to figure out that for any path 
different from the path shown with heavy solid line in Fig.4c, the 
particle will make a longer excursion to arrive at site 2. 

\begin{figure}
\resizebox{\textwidth}{!}{\includegraphics{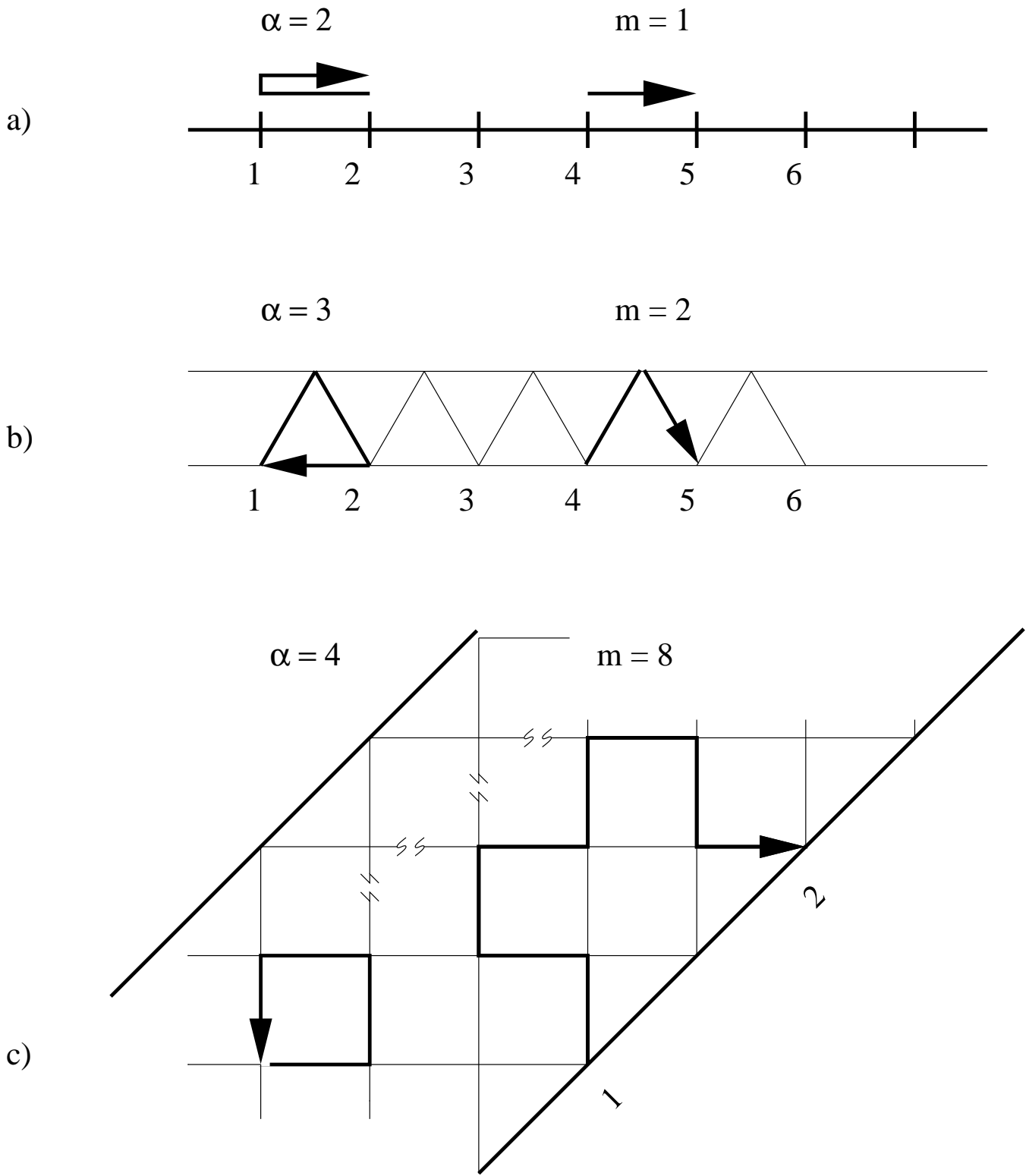}}
\caption{Propagation channels in one- and two-dimensional lattices. 
Heavy lines with arrows show particle displacements in elementary loop 
($\alpha=2,\,3,\,4$) and minimal trajectory between two consecutive sites 
on the propagation line ($m=1,\,2,\,8$). Propagation in the 
one-dimensional lattice (a) and in the triangular lattice (b) is
discussed in \cite{grosfils}; (c) refers to Langton's ant 
(see \cite{ant} and \cite{boon}).}
\end{figure}

With the specification of the time delays in \Eq{tau}, we can give an 
explicit formulation of the quantities which appear in the continuous 
limit equation. The average displacement time and the variance read
respectively
\beqa{ap2}
\la\tau\ra\, =\, \sum_{j=0}^n\,(1\,+\,\alpha j)\,p_j\,m\,\delta t
%&=& \sum_{j=0}^n\,p_j\,\delta t\,+\,\alpha\,\sum_{j=0}^n j\,p_j\,\delta t\nm
\,=\, (1\,+\,\alpha\,\la j\ra)\,m\,\delta t \;,
\eeqa
and 
\beqa{ap3}
\la \tau^2\ra -\la \tau\ra ^2&=&\left\{\,\sum_{j=0}^n\,(1+\alpha j)^2 
p_j\,-\,
[\sum_{j=0}^n(1\,+\,\alpha\,j\,)\,p_j]^2\right\}\,m^2 (\delta t)^2 \nm
%&=&\{ \sum_{j=0}^n\,p_j\, +\, 2\alpha\,\sum_{j=0}^n j\,p_j\,+\, \alpha^2
%\sum_{j=0}^n j^2\,p_j\, -\,[1+\alpha\,\sum_{j=0}^n j\,p_j]^2\}\,m^2 
&=&\left\{\sum_{j=0}^n j^2\,p_j\,-[\sum_{j=0}^n j\,p_j]^2\right\}\,
\alpha^2\,m^2 (\delta t)^2 \nm
&=&(\la j^2\ra -\la j\ra ^2)\,\alpha^2\,m^2 (\delta t)^2 \;,
\eeqa
where from it follows that
\beq{ap6}
\frac{1}{c}\,=\,
(1+\alpha\la j\ra)\,m\,\frac{\delta t}{\delta r}\;,
\eeq
and
\beq{ap7}
\gamma\,=\,
(\la j^2\ra \,-\,\la j\ra ^2)\,(\alpha\,m)^2 \, \frac{(\delta t)^2}
{\delta r}\;.
\eeq
The value of $c$ and of $\gamma$ depends on the spin orientation 
probability $q$, the probability that a site be in the $\uparrow$ state
at the initial time.
The derivation of the propagation-dispersion equation proceeds exactly
along the lines of Section \ref{P-D_Eq} and yields Eq.\Eq{a19} with
$c$ and $\gamma$ given by \Eq{ap6} and \Eq{ap7} respectively.
Figure 5 shows the agreement of the analytical results with the simulation 
data for the 1-D and 2-D lattices. For instance, for the 1-D
spin-lattice whose dynamics is described in the introductory section,
$m=1, \alpha = 2$ (see Fig.4a), $j=0,1$, and 
$\la j \ra = \sum_j j p_j = 1-q$ and $\la j^2 \ra - \la j \ra^2 =
\sum_j j^2 p_j -  (\sum_j j p_j)^2 = q(1-q)$; so $c=(3-2q)^{-1}$ and
$\gamma = 4q(1-q)$. The corresponding numerical values are given
in the figure caption for both the 1-D lattice (Fig.5a) and the
2-D triangular lattice (Fig.5b).

In a recent paper \cite{buni_khlabys}, Bunimovich and Khlabystova,
referring to a preliminary version of the present work \cite{archives},
obtain the same propagation-dispersion equation for the case of a
particle moving in a rigid environment \cite{bunimov}: the rigidity
factor  $\tilde r$ is defined as the number of visits of the particle
to a site necessary to flip its state ($\uparrow \Longleftrightarrow 
\downarrow$). The time delay probabilities are then space-dependent.
For odd rigidity (NOS model in \cite{buni_khlabys}), there is no 
qualitative difference for $c$ and $\gamma$ with the simple 1-D case 
($\tilde r =1$) discussed above (compare the expressions  given in the
previous paragraph with those in Section 4.1 of \cite{buni_khlabys}). 
However when the rigidity factor takes even values, the propagation 
speed and the dispersion coefficient become space-dependent 
(see Section 4.2 in \cite{buni_khlabys}). So the microscopic 
dynamics of the (NOS) model with even rigidity offers an example 
where the continuous limit is given by the generalized 
propagation-dispersion equation \Eq{geneq}. 

\begin{figure}
\resizebox{\textwidth}{!}{\includegraphics{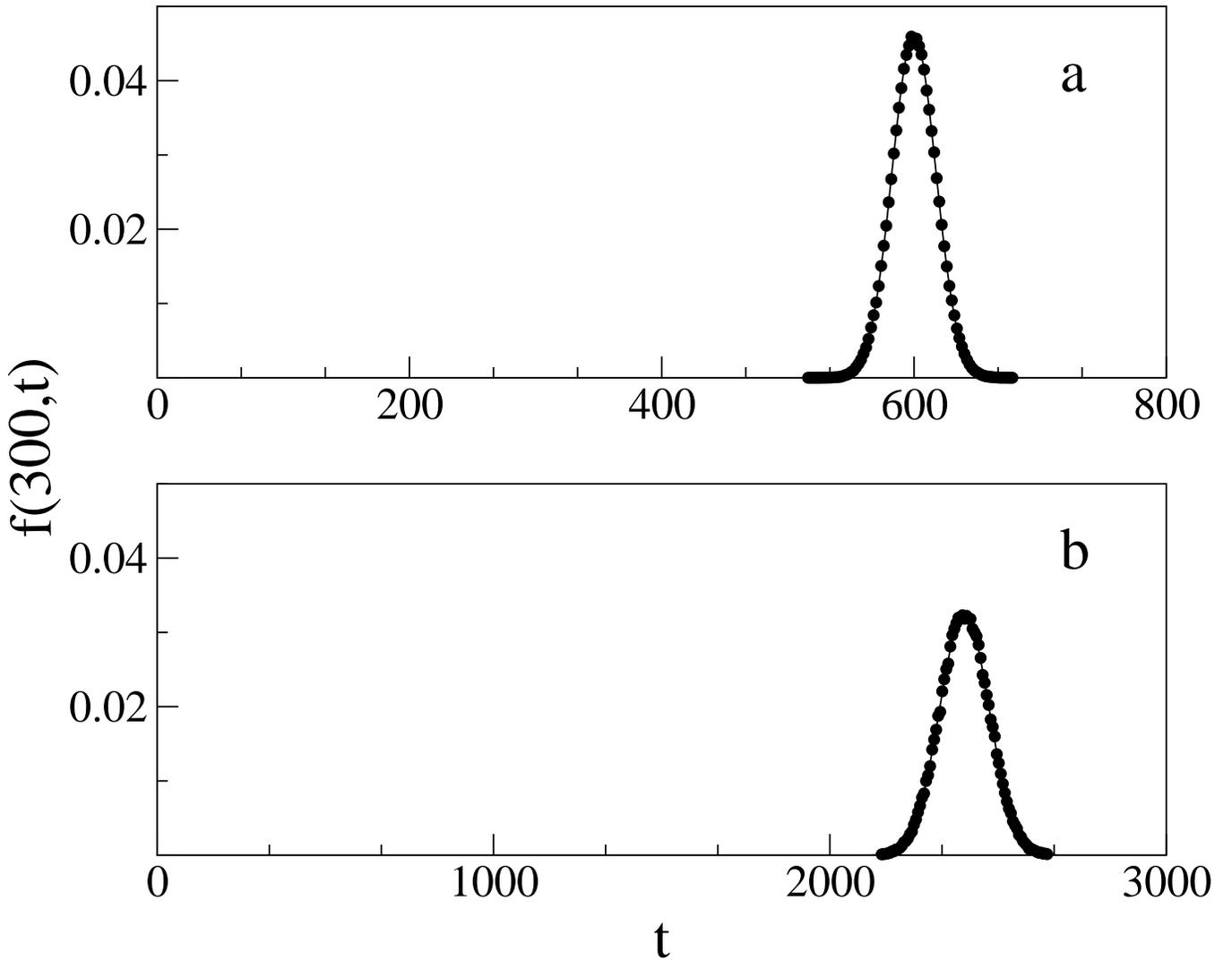}}
\caption{Probability distribution $f(r=300,\,t)$ for
propagation in $1-$D and $2-$D lattices. Comparison between
numerical simulation results (dots) and theory, Eq.\Eq{a20} (solid line).
(a) $1-$D lattice (see Fig.2a) with $q=0.5$: $c=(3-2\,q)^{-1} = 0.5$ and
$\gamma = 4q(1-q) = 1$; half-width $ =\sqrt {2 \gamma r}
= 10\sqrt6 \simeq 24.5$.
(b) $2-$D triangular lattice (see Fig.2b) with $q=0.5$: $c=1/8$ and
$\gamma = 72 q(1-q) = 18$ (see \cite{grosfils});  half-width $ 
=\sqrt {2 \gamma r} = 60\sqrt3  \simeq 104$.} 
\end{figure}

The explicit expression \Eq{tau} for $\tau_j$ based on the automaton 
dynamics yields analytical expressions for $c$ and $\gamma$ in terms 
of the spin-lattice characteristics and there from in terms of the 
probability $q$. Concomitantly there is an explicit reference to a 
feed-back mechanism where the dynamics modifies locally the substrate 
which in turn modifies the dynamics, and there are systems where 
this mechanism should be important. However we emphasize that the 
specification \Eq{tau} of the delay time $\tau_j$ is not indispensable. 
The propagation-dispersion equation \Eq{a19} is general as it follows from 
the continuous limit of the first visit equation \Eq{a0} without 
recourse to \Eq{tau} as shown in Section \ref{P-D_Eq}. It
suffices that there exists a distribution of time delays with 
$\sum_{j=0}^n\,p_j=1$ and with finite moments $\la \tau^a\ra$, 
to obtain \Eq{a19} from \Eq{a0}. This establishes the validity of the
propagation-dispersion equation for a class of systems whose dynamics
is subject to time-delays independently of the
underlying microscopic mechanism responsible for the delays.

%%\bigskip

\end{document}